\RequirePackage{fix-cm}
\documentclass[twocolumn,a4paper,natbib]{svjour3}          
\usepackage{graphicx}
\journalname{Journal of Geodesy}
\begin{document}
\title{The AuScope Geodetic VLBI Array}


\author{J.~E.~J.~Lovell \and
J.~N.~McCallum \and
P.~B.~Reid \and
P.~M.~McCulloch \and
B.~E.~Baynes \and
J.~M.~Dickey \and
S.~S.~Shabala \and
C.~S.~Watson \and
O.~Titov \and
R.~Ruddick \and
R.~Twilley \and
C.~Reynolds \and
S.~J.~Tingay \and
P.~Shield \and
R.~Adada \and
S.~P.~Ellingsen \and
J.~S.~Morgan \and
H.~E.~Bignall
}

\authorrunning{Lovell et al.} 

\institute{J. E. J. Lovell \at
	School of Mathematics and Physics, University of Tasmania, Private Bag 37, Hobart, 7001, Australia \\
	Tel.: +61-3-6226-7256\\
	Fax: +61-3-6226-2410\\
	\email{Jim.Lovell@utas.edu.au}
	\and
	J. N. McCallum \at
	School of Mathematics and Physics, University of Tasmania, Private Bag 37, Hobart, 7001, Australia \\
	Tel.: +61-3-6226-7529\\
	Fax: +61-3-6226-2410\\
	\email{Jamie.McCallum@utas.edu.au}
	\and
	P. B. Reid \at
	School of Mathematics and Physics, University of Tasmania, Private Bag 37, Hobart, 7001, Australia \\
	Tel.: +61-3-6226-6383\\
	Fax: +61-3-6226-2410\\
	\email{Brett. Reid@utas.edu.au}
	\and
	P. M. McCulloch \at
	School of Mathematics and Physics, University of Tasmania, Private Bag 37, Hobart, 7001, Australia \\
	Tel.: +61-3-6226-2420\\
	Fax: +61-3-6226-2410\\
	\email{Peter.McCulloch@iinet.net.au}
	\and
	B. E. Baynes \at
	School of Mathematics and Physics, University of Tasmania, Private Bag 37, Hobart, 7001, Australia \\
	Tel.: +61-3-6226-6383\\
	Fax: +61-3-6226-2410\\
	\email{Eric.Baynes@utas.edu.au}
	\and
	J. M. Dickey \at
	School of Mathematics and Physics, University of Tasmania, Private Bag 37, Hobart, 7001, Australia \\
	Tel.: +61-3-6226-2447\\
	Fax: +61-3-6226-2410\\
	\email{John.Dickey@utas.edu.au}
	\and
	S. S. Shabala \at
	School of Mathematics and Physics, University of Tasmania, Private Bag 37, Hobart, 7001, Australia \\
	Tel.: +61-3-6226-8502\\
	Fax: +61-3-6226-2410\\
	\email{Stanislav.Shabala@utas.edu.au}
	\and
	C. S. Watson \at
	Surveying and Spatial Science Group, School of Geography and Environmental Studies, University of Tasmania, Private Bag 76, Hobart, 7001, Australia \\
	Tel.: +61-3-6226-2489\\
	Fax: +61-3-6226-7628\\
	\email{Christopher.Watson@utas.edu.au}
	\and
	O. Titov \at
	Geoscience Australia, P.O. Box 378, Canberra, ACT 2601, Australia\\
	Tel.: +61-2-6249-9064\\
	Fax: +61-2-6249-9929\\
	\email{Oleg.Titov@ga.gov.au}
	\and
	R. Ruddick \at
	Geoscience Australia, P.O. Box 378, Canberra, ACT 2601, Australia\\
	Tel.: +61-2-6249-9426\\
	Fax: +61-2-6249-9929\\
	\email{Ryan.Ruddick@ga.gov.au}
	\and
	R. Twilley \at
	Geoscience Australia, P.O. Box 378, Canberra, ACT 2601, Australia\\
	Tel.: +61-2-6249-9066\\
	Fax: +61-2-6249-9929\\
	\email{Bob.Twilley@ga.gov.au}
	\and
	C. Reynolds \at
	International Centre for Radio Astronomy Research, Curtin University, GPO Box U1987, Perth, WA 6845, Australia\\
	Tel.: +61-8-9266-3785\\
	Fax: +61-8-9266-9246\\
	\email{c.reynolds@curtin.edu.au}
	\and
	S. J. Tingay \at
	International Centre for Radio Astronomy Research, Curtin University, GPO Box U1987, Perth, WA 6845, Australia\\
	Tel.: +61-8-9266-3516\\
	Fax: +61-8-9266-9246\\
	\email{Steven.Tingay@icrar.org}
	\and
	P. Shield \at
	InterTronic Solutions Inc., 452, Aime-Vincent, Vaudreuil-Dorion, Quebec, Canada, J7V 5V5 \\
	Tel.: +1-450-424-5666\\
	Fax: +1-450-424-6611\\
	\email{pshield@intertronicsolutions.com}
	\and
	R. Adada \at
	Cobham Antenna Systems, SATCOM Land, 1551 College Park Business Center, Orlando, FL 32804, USA\\
	Tel.: +1-407-650-9054 ext.101\\
	Fax: +1-407-650-9086\\
	\email{Rami.Adada@cobham.com} 
	\and
	S. P. Ellingsen \at
	School of Mathematics and Physics, University of Tasmania, Private Bag 37, Hobart, 7001, Australia \\
	Tel.: +61-3-6226-7588\\
	Fax: +61-3-6226-2410\\
	\email{Simon.Ellingsen@utas.edu.au}
	\and
	J. S. Morgan \at
	International Centre for Radio Astronomy Research, Curtin University, GPO Box U1987, Perth, WA 6845, Australia\\
	Tel.: +61-8-9266-9105\\
	Fax: +61-8-9266-9246\\
	\email{john.morgan@curtin.edu.au}
	\and
	H. E. Bignall \at
	International Centre for Radio Astronomy Research, Curtin University, GPO Box U1987, Perth, WA 6845, Australia\\
	Tel.: +61-8-9266-9245\\
	Fax: +61-8-9266-9246\\
	\email{h.bignall@curtin.edu.au}
}


\date{Received: 22 August 2012 / Accepted: 18 February 2013}

\maketitle

\begin{abstract}
The AuScope geodetic Very Long Baseline Interferometry array consists of three new 12~m radio telescopes and a correlation facility in Australia. The telescopes at Hobart (Tasmania), Katherine (Northern Territory) and Yarragadee (Western Australia) are co-located with other space geodetic techniques including Global Navigation Satellite Systems (GNSS) and gravity infrastructure, and in the case of Yarragadee, Satellite Laser Ranging (SLR) and Doppler Orbitography and Radiopositioning Integrated by Satellite (DORIS) facilities. The correlation facility is based in Perth (Western Australia).

This new facility will make significant contributions to improving the densification of the International Celestial Reference Frame in the Southern Hemisphere, and subsequently enhance the International Terrestrial Reference Frame through the ability to detect and mitigate systematic error. This, combined with the simultaneous densification of the GNSS network across Australia will enable the improved measurement of intraplate deformation across the Australian tectonic plate.

In this paper we present a description of this new infrastructure and present some initial results, including telescope performance measurements and positions of the telescopes in the International Terrestrial Reference Frame. We show that this array is already capable of achieving centimetre precision over typical long-baselines and that network and reference source systematic effects must be further improved to reach the ambitious goals of VLBI2010.

\keywords{Geodesy \and Very Long Baseline Interferometry (VLBI) \and Celestial Reference Frame (CRF) \and Terrestrial Reference Frame (TRF)}
\end{abstract}

\section{Introduction and Background}
\label{intro}

The AuScope Very Long Baseline Interferometry (VLBI) array was
constructed as part of the National Cooperative Research
Infrastructure Strategy (NCRIS) funded by the Australian Department of
Innovation, Industry, Science, and Research \citep{NCRIS_web}.  The
VLBI array is one part of AuScope, a diverse framework of infrastructure for
research in geological, geochemical, geophysical, and geospatial
subjects \citep{AuScope_web}.  The geospatial component of AuScope is
composed of four kinds of infrastructure: the VLBI array, a suite of
upgrades to Satellite Laser Ranging (SLR) facilities, a continent-wide array of $\sim100$
Global Navigation Satellite Systems (GNSS) ground stations, and a
number of precision gravity measurement instruments.  These facilities
work together, and in cooperation with similar equipment throughout
the world, to define and monitor the geodetic reference frame and
enable the improved geophysical interpretation of space geodetic data.

Various reference systems may be defined to form the foundation to
realise a Terrestrial Reference Frame (TRF) which can be used to
express positions on the Earth's dynamic surface
\citep{petit2010}. 
The space geodetic
technique of VLBI provides a vital contribution to the realisation of
the International TRF \citep[ITRF2008,][]{2011JGeod..85..457A}, particularly
through its ability to define the scale of the frame and its
orientation with respect to the International Celestial Reference Frame 
\citep[ICRF2,][]{fey2009}.  Observations with the VLBI technique uniquely define the inertial ICRF and continue to maintain and improve its precision. 

The Earth Orientation Parameters (EOPs) are used to link the ICRF and ITRF and comprise five components: two nutation angles, two of polar motion and Universal Time (UT1). While other space-geodetic techniques (e.g. GNSS, SLR) provide information on polar motion, VLBI has capability of measuring all five EOPs and is unique in its ability to obtain the nutation angles and UT1. Therefore the ICRF2 as a realisation of the International Celestial Reference System (ICRS) must be carefully supported and maintained. See \citet{schuhandbehrend} for a review of the VLBI technique and its important contribution to geodesy. 

Precise data on nutation is important in its own
right as a probe of angular momentum changes in the atmosphere,
oceans, and interior of the Earth \citep[e.g.][]{Herring_etal_2002}.
Importantly, other space geodetic techniques including GNSS that densify the ITRF, depend upon
frequent updates to the EOPs, in particular UT1 and precession/nutation, as determined through VLBI
observations. These estimates are used a-priori and are often tightly
constrained in order to achieve precise positioning over the
Earth. Combined, this is the mission of the International VLBI Service
(IVS) for Geodesy and Astrometry \citep{schuhandbehrend}.

\begin{figure*}
\includegraphics[width=1.0\textwidth]{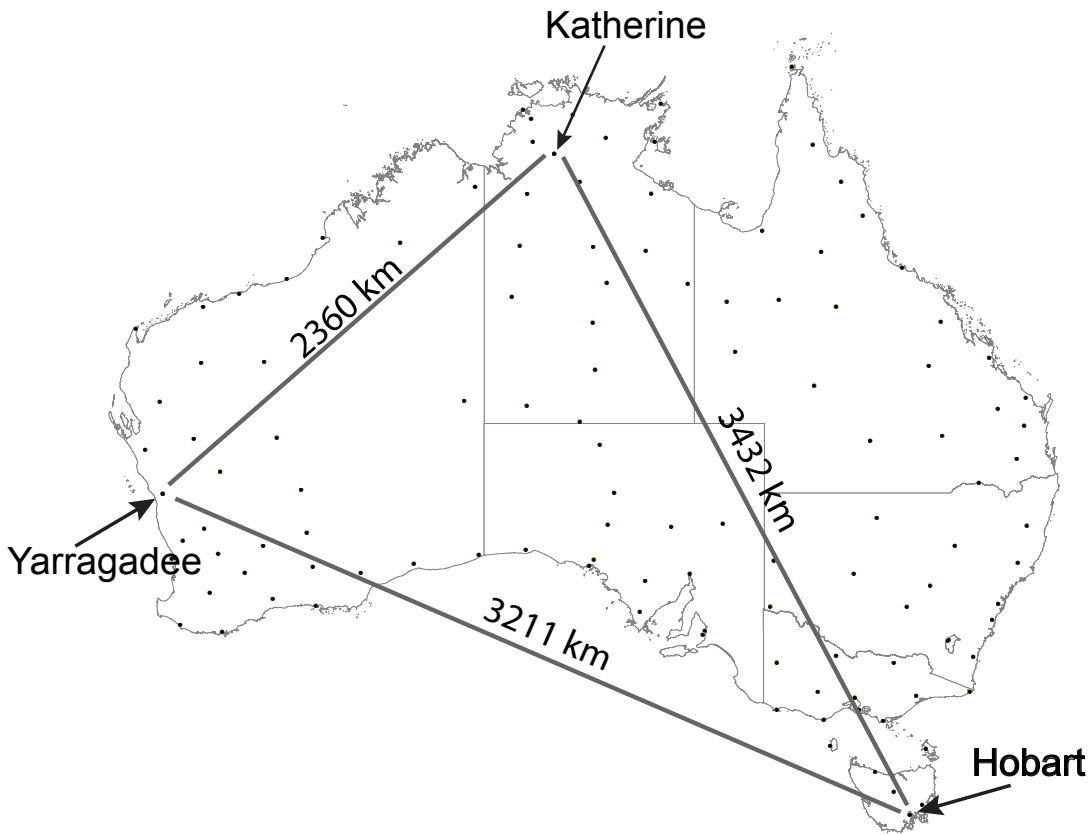}
\caption{The geographical distribution of VLBI and GNSS infrastructure for AuScope. The locations of the new 12\,m telescopes are labelled and the new GNSS sites are indicated by filled dots.}
\label{fig:auscope_array}
\end{figure*}

The 26\,m telescope at the Mt. Pleasant Observatory of the University
of Tasmania, near Hobart, has been a member of the IVS network and its
predecessors for more than 20 years.  Hobart is one of the few IVS
network sites in the Southern Hemisphere that can work all year round
and is able to contribute its data rapidly to the IVS correlation
centres in North America and Europe.  As of 2006, the only IVS
observatories at mid to high southern latitudes capable of frequent,
year-round observations were in Chile (the Transportable Integrated Geodetic Observatory (TIGO)), South Africa
(Hartebeesthoek Radio Astronomy Observatory), and Hobart.  In order to
expand the ability of the IVS to monitor ICRF sources south of
declination $-40\,^{\circ}$, it was very important to increase the
number of telescopes at southern latitudes \citep{titov2009}.  Thus
one of the major objectives of the AuScope infrastructure strategy was
to build three new VLBI telescopes in Australia.  
The physical positioning of the telescopes
seeks to optimise the possible improvements to the ICRF and ITRF from
the Australian continent (Figure~\ref{fig:auscope_array}). The GNSS
array then provides the densification required to achieve the goal of
enabling high precision coordination across the continent.

The three new AuScope VLBI antennas are located at Hobart (Tasmania),
Katherine (Northern Territory) and Yarragadee (Western Australia) (Figure~\ref{fig:auscope_array}).
The new Hobart telescope is co-located with the existing 26\,m
telescope to preserve the more than 20 year VLBI time series at the
site. Midway between the 26\,m and 12\,m telescopes is the HOB2 GNSS
installation which has been a core site of the International GNSS
Service (IGS) since its inception. A hut capable of housing a mobile
gravimeter is also co-located on the site. The Yarragadee telescope
provides a far western point on the continent and is co-located with
multiple existing geodetic techniques including SLR, GNSS, Doppler Orbitography and Radiopositioning Integrated by Satellite (DORIS) and
gravity. The Katherine site is new and provides a central longitude,
northern site. The telescope is co-located with a new GNSS site that
forms part of the AuScope GNSS network.

The design of the new AuScope telescopes was driven by a new paradigm
for the IVS, defined by a study team in the early 2000's
and given the name VLBI2010 \citep{Niell_etal_2007,petra2010}.  A critical
specification for the new telescope design is drive speed.  This
provides better performance for the specific needs of a geodetic VLBI
array because it allows the telescopes to observe more sources at
wide separations on the sky in a given time interval
\citep{titov2007}.  Rapid
switching from source to source is a critical telescope design parameter in order to solve for the effects of the troposphere on the propagation delay of the radio signals.  With this in mind, the VLBI2010 model favours small
telescopes (12\,m rather than $\sim$25\,m diameter) with powerful
drive motors.

\begin{sloppypar}
Each AuScope VLBI observatory is equipped with a 12.1\,m diameter main
reflector designed and constructed by COBHAM Satcom, Patriot Products
division. All sites are equipped with dual polarization S and X-band
($2.2-2.4$~GHz and $8.1-9.1$~GHz respectively) feeds from COBHAM with room temperature receivers developed at the
University of Tasmania by Prof. Peter McCulloch.  Data digitisation
and formatting is managed by the Digital Base Band Converter (DBBC)
system from HAT-Lab, and data are recorded using the Conduant Mark5B+
system. Each site is equipped with VCH-1005A Hydrogen maser time and
frequency standards from Vremya-CH.
\end{sloppypar}

The AuScope infrastructure funding also included an allocation to establish a software correlator, the Curtin
University Parallel Processor for Astronomy (CUPPA), a 20 node beowulf
compute cluster running the DiFX software correlator
\citep{deller2011} to process data from the array.

In this paper we provide a detailed description of the telescopes and
associated hardware, signal processing, control and monitoring
equipment and the correlation facility. Lastly, we describe the
performance of the telescopes and present some initial geodetic
results from the array.

\section{The Observatories}
\label{obs}

\subsection{Antenna Characteristics}
\label{antenna}

\begin{figure*}
\includegraphics[width=0.5\textwidth]{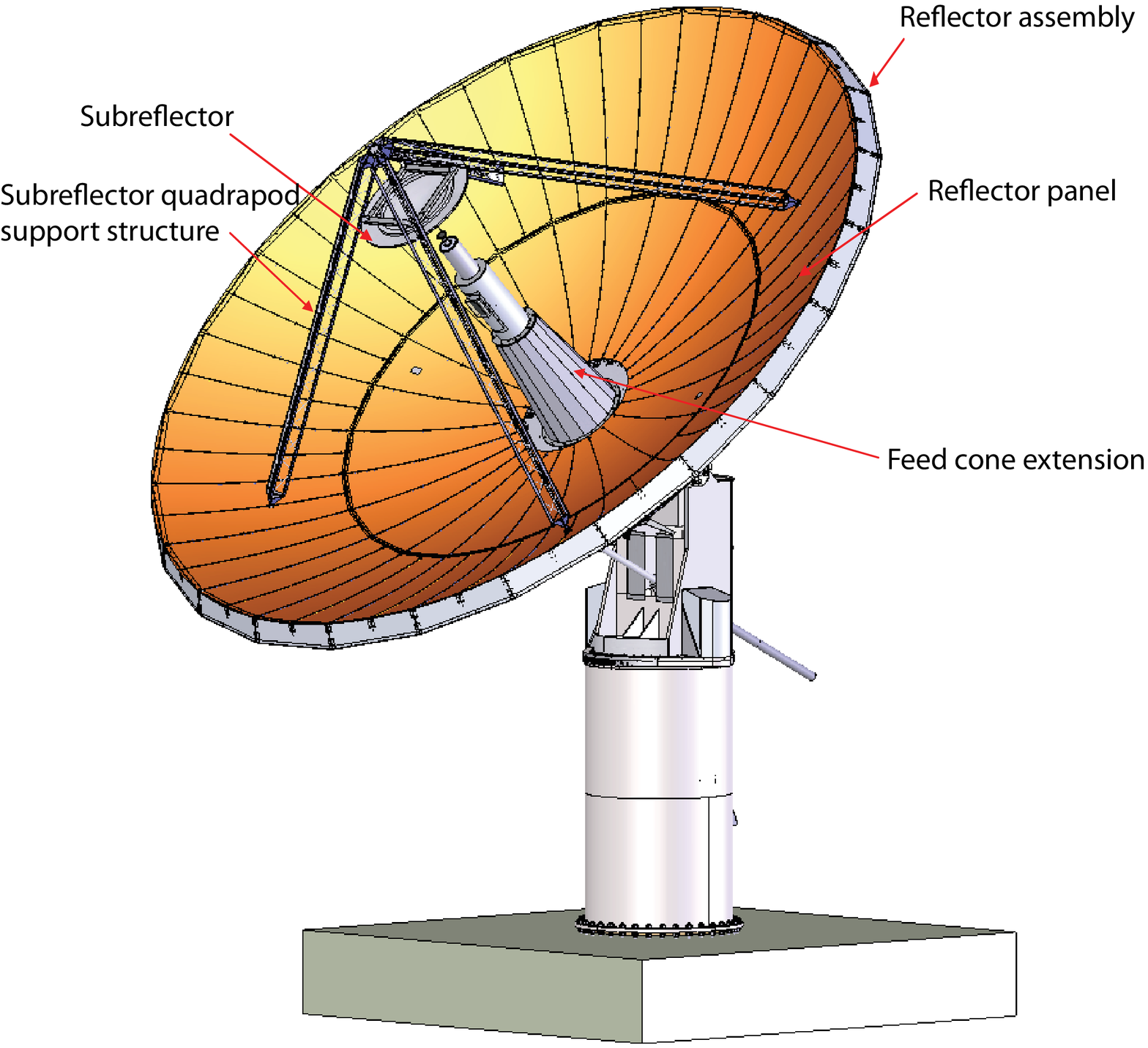}
\includegraphics[width=0.5\textwidth]{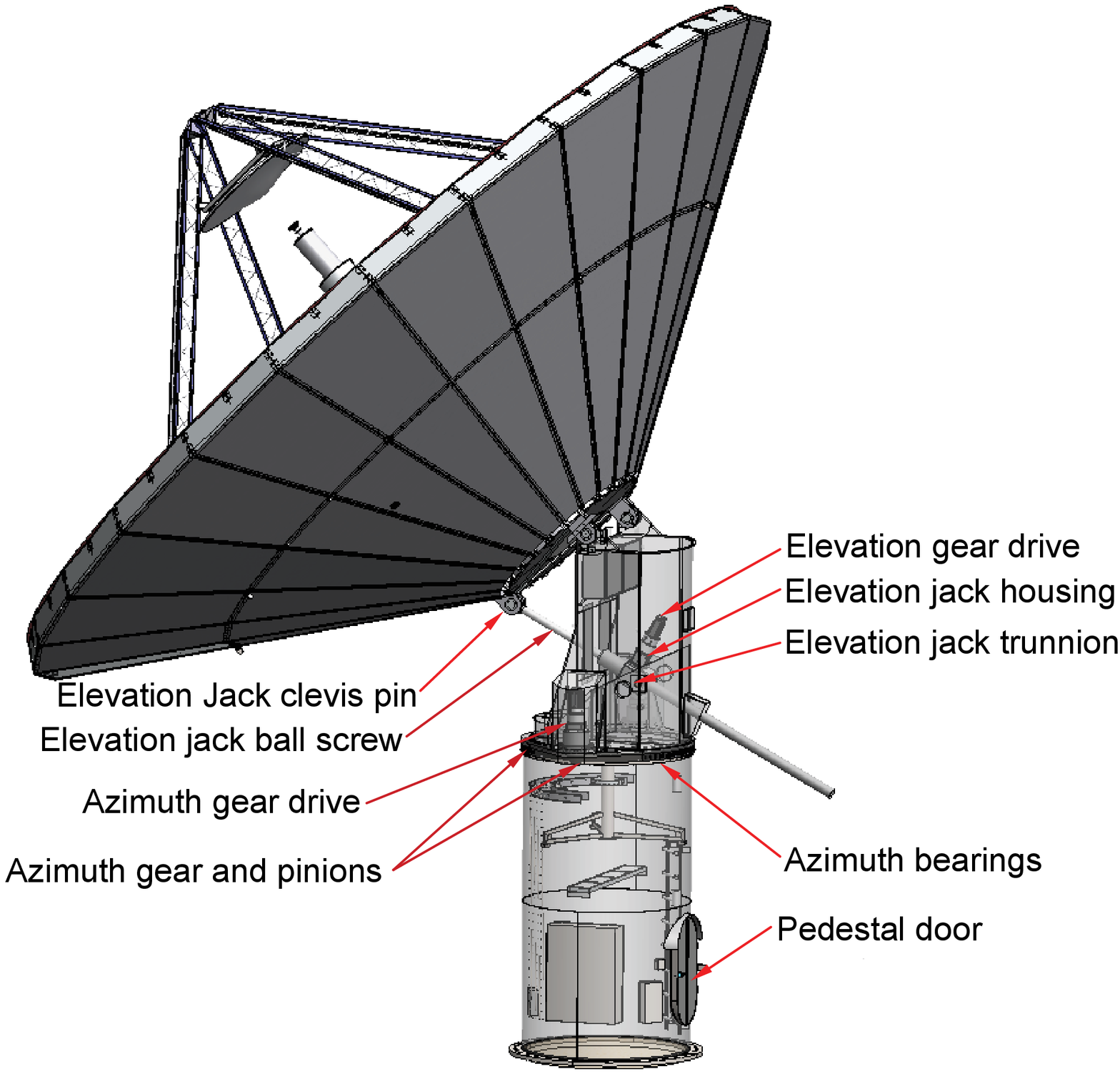}
\caption{Drawings of the antennas used in the AuScope VLBI array with the main components labeled.}
\label{fig:antennas}       
\end{figure*}

The 12\,m cassegrain shaped reflector, pedestal mounted antennas
(Figure~\ref{fig:antennas}) are of a unique design in this size range
as they use a large ball jack screw permanently in compression for
elevation. This elevation drive therefore uses just one motor as
backlash is taken care of by keeping the jack in compression. For
azimuth however two motors and pinions are used on the main azimuth
gear/bearing and the motors are electrically torque biased against
each other to eliminate any backlash. The motors are high torque
standard AC brushless servo motors and two stages of gearbox reduction
are used on both axes to increase the torque and reduce speed. The
main azimuth gear/bearing is a massively oversized, nearly 2.3\,m
diameter, opposing roller bearing system ensuring great precision and
long life.

The combination of materials used in the antenna is somewhat
unconventional. The primary reflector, including its shape-determining
support ribs, hub etc are aluminium but below the elevation bearings
the whole turning head and pedestal are made of steel,
and painted for thermal and corrosion prevention. The subreflector is
a custom shaped carbon fiber reinforced polymer moulding.

The main 12\,m reflector is fabricated with water jet cut radial ribs
that determine the final surface 0.38\,mm RMS shape, a rigid aluminium
hub, stretch formed precision front reflector panels and rear close
out panels. The stretch formed panels need to be attached to the ribs
to hold their shape, and the rear close out panels strengthen the
radial ribs. Once the main reflector is fully assembled and locked up
tight it is a strong structure. However, there is essentially no
individual panel surface adjustability.

The antenna can slew at up to 5 deg/s in azimuth and 1.5 deg/s in elevation. 
The control system monitors and controls the whole system using 26 bit
optical absolute position feedback devices mounted right on the axes
of motion. The RFI-tight control system utilises standard modern COTS
servo systems and components. The control software was developed under
contract by Dr.\,Mark Godwin in the UK.

\subsection{Telescope optics and feed}
\label{feed}
The COBHAM Satcom 12\,m antenna system was jointly developed with
NASA's Jet Propulsion Laboratory as part of the proof of concept for
the Deep Space NetworkÕs large array proposal \citep{gatti}. The radio
frequency optical design of the system 
\citep[based on the work presented in][]{Imbriale} utilises a shaped Cassegrain configuration optimised
for maximum $G/T$ performance with a 1.8\,m subreflector.

\begin{figure*}
\includegraphics[width=\textwidth]{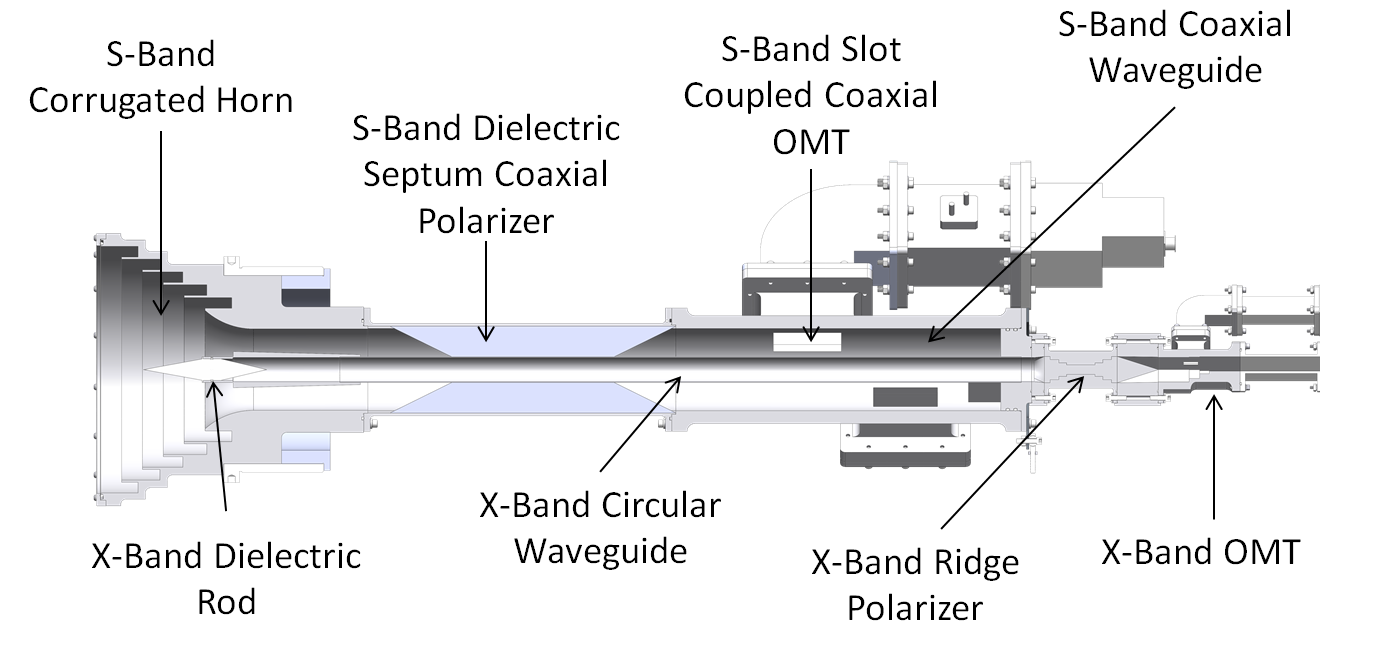}
\caption{Online Resource 1: A cross-sectional diagram of the coaxial S/X feed developed for the 12\,m antennas.}
\label{fig:or1}
\end{figure*}

The feed for the antenna system, developed by COBHAM in response to
interest from the radio astronomy community, employs a 4-port coaxial
configuration to allow for simultaneous reception of right- and
left-hand circular polarisation signals (RCP and LCP respectively) at
both the S and X bands. The feedÕs operating frequency ranges are
2.2--2.4\,GHz at S-Band and 8.1--9.1\,GHz at X-Band.  A cross section
diagram of the feed is presented in Online Resource 1 (Figure~\ref{fig:or1}). A breakdown of
the expected system performance is tabulated in Online Resource 2 (Table~\ref{tab:or2}) but
can be summarised by System Equivalent Flux Densities (SEFDs) of
$3500$~Jy and $3400$~Jy at S- and X-bands respectively.

\begin{table*}
\caption{Online Resource 2: Expected Performance of the 12m Radio Telescope with Dual Band S/X Feed}
\label{tab:or2}
\begin{tabular}{|p{1.5 cm}|p{3 cm}|c|c|p{3 cm}|}
\hline
 & {\bf Parameter} & {\bf S Band} & {\bf X Band} & {\bf Additional Notes} \\\hline
{\bf Losses} &Spillover \& Illumination loss (dB)	& 0.81 & 0.66	& i.e. 83\% and 86\% efficiency respectively\\ \cline{2-5}
           & Blockage Loss (dB)                    & 0.71 & 0.71 & Blocked aperture due to Subreflector and Quadripod \\ \cline{2-5}
           & VSWR Loss (dB)                        & 0.14 & 0.04	& $<-15$ dB and $-20$ dB Return Loss\\ \cline{2-5}
           & Port to Port Isolation (dB)            & 0.04 & 0.04 & $>20$ dB isolation \\ \cline{2-5}
           & Cross-polarisation loss (dB)                         & 0.03 & 0.03  & Axial Ratio $< 1.5$ dB\\ \cline{2-5}
           & Surface RMS Loss (dB)	              & 0.01 & 0.09  & Using Ruze equation and RMS of 0.38 mm \\ \cline{2-5}
           & Ohmic Losses (dB)                     & 0.5  & 0.35   & In Window, Horn, Polarizer and OMT\\ \cline{2-5}
           & Total Loss (dB)                           & 2.24 & 1.92	 & \\ \cline{2-5}
           & Total Aperture Efficiency             & 59.7\%	& 64.3\%	 & \\ \hline
{\bf Noise Temperature}	& LNA (K)                     & 30      & 50   & Manufacturer Specification \\ \cline{2-5}
	                     & Feed (K)                    & 35 & 25 & Based on Ohmic losses above at 290 K\\ \cline{2-5}
                            & Quadripod, Subreflector and S-Band Horn Scatter (K) & 10 & 5 & \\ \cline{2-5}
                            & Galactic + Atmospheric (K) & 10 & 10 & \\ \cline{2-5}
                            & System Noise temperature (K) & 85	 & 90 & \\ \hline
{\bf Sensitivity} & SEFD (Jy) &  3500 & 3400 & \\ \hline
\end{tabular}
\end{table*}

\subsection{Front end receiver (feed tube and hub)}
\label{front_end}
\begin{figure}
\includegraphics[width=0.5\textwidth]{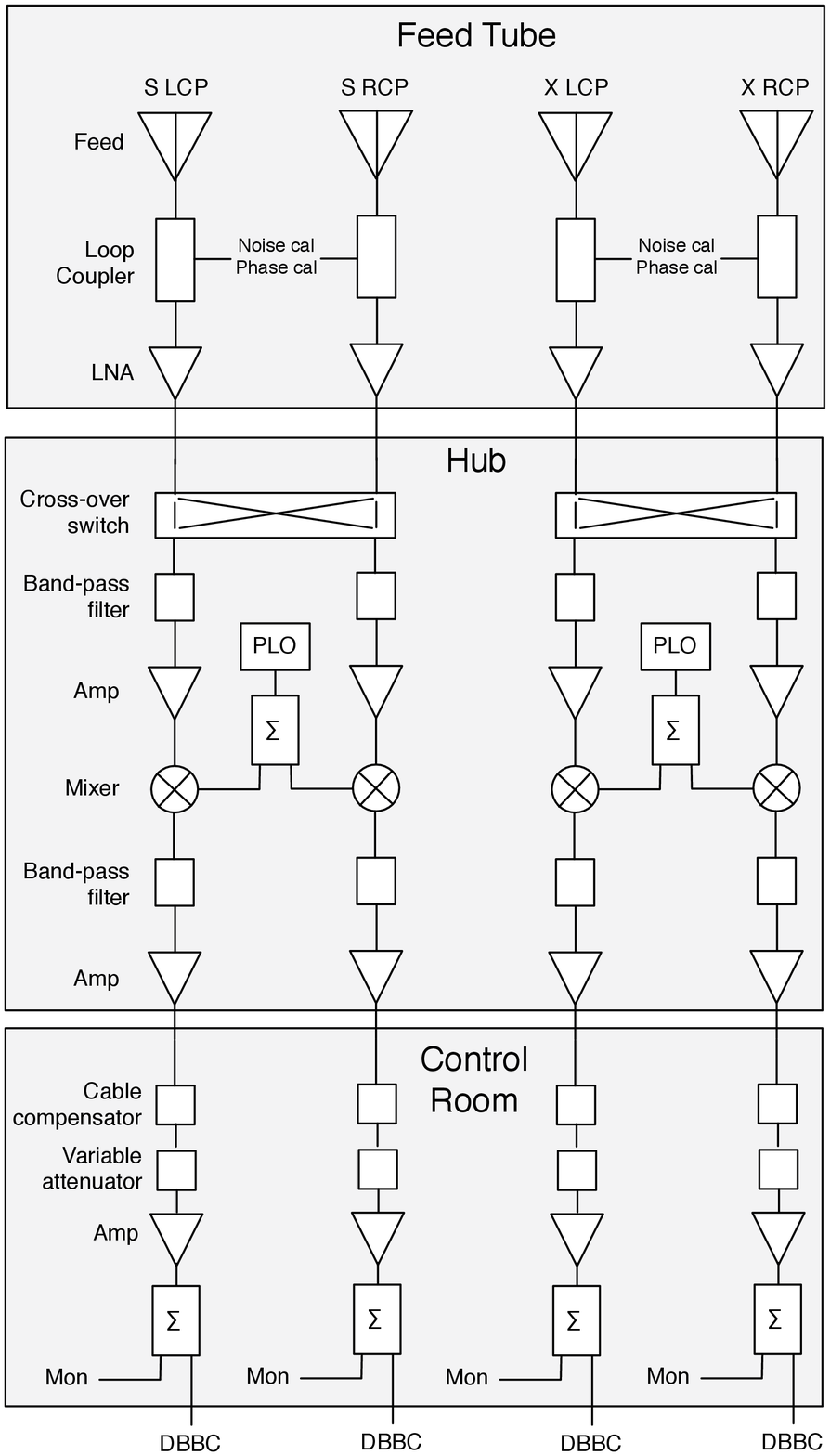}
\caption{A schematic, illustrating the signal path from the feed at the focal point to the receiver in the antenna hub (immediately below the feed cone extension labeled in Figure~\ref{fig:antennas}) to the IF signal outputs to the DBBC.}
\label{fig:rx}     
\end{figure}

The receiver takes the dual circular polarisation signals from Low
Noise Amplifiers (LNA) on the feed, each with $\sim35$\,dB of gain,
and down-converts them to baseband suitable for detection and further
processing. Figure~\ref{fig:rx} provides an overview of the signal
path, a description of which follows in this and the next two Subsections.

The receiver allows for injection of calibration signals from noise
diodes, a tone generator and/or a separate phase-calibration pulse
generator. Two noise diode signals may be injected: a low power signal
($\sim 0.1$ dB) suitable for modulation to enable continuous system
temperature (Tsys) measurements; or from a higher-powered signal
($\sim 0.5$ dB) for single Tsys measurements or system testing. A tone
injection port is available for coherence testing or other maintenance
and configuration tasks.  

In the
situation where only a single polarisation is required (as is common
in current geodetic VLBI observations or spacecraft tracking
applications) some redundancy is provided by cross-over switches in
the signal path to permit quick swapping of polarisations. This allows
for continued operation in the event of a component failure downstream
of the switches.

The receiver has front-end filters which
pass through 2.0 to 2.5 GHz at S-band and 8.0 to 9.2 GHz at X-band.
The receiver system also permits inclusion of notch
filters to block interference signals prior to down-conversion. 

The front-end receiver down-converts the S- and X-band signals with local oscillators at 1900~MHz and 7600~MHz respectively, phase-locked to a 100~MHz tone from the hydrogen maser.

\subsection{Back end receiver and recorder (control room)}
\label{back_end}

The down-converted signals from the front end receiver are then carried underground to the control room where 
an Intermediate Frequency (IF) unit provides some signal
conditioning to bring output power levels into range, then into the DBBC for sampling. The sampled
data are then recorded to disk using a Mark5B+ unit. 

The IF unit consists of 4 signal chains (RCP and LCP of S- and X-band),
each with tuned cable compensators, IF amplifiers and independent
controllable attenuator arrays. The outputs of each IF are split with
two outputs going to the inputs of the DBBC conditioning modules,
together with one other output as an analogue monitor point. In normal
operation, these monitoring points are connected to a remotely
controlled selector device which determines the input to a broad-band
power sensor. A network-accessible spectrum analyser is also connected
to this analogue monitor point. This is particularly useful for
detecting the presence of interfering signals.

\begin{sloppypar}
The DBBC system is composed of a series of modules which filter and
sample the input IFs and produce a number of output baseband-converted
channels. The IF selection and bandpass filtering is handled by the Conditioning
Modules. These modules support up to 4 input IFs, and the selected IF
can then be filtered by one of three selectable, 512~MHz wide, bandpass filters (in
the ranges of 0--512, 512--1024 and 1024--1536 MHz).
The total power level of the IF
is then adjusted using variable attenuators within the Conditioning
Modules before being sampled. The DBBC sampler is synchronised to the
station maser through a 10 MHz and 1 pulse-per-second (1PPS)
signal. There are four tuneable baseband converter units supplied from
each of the 4 Conditioning Modules, which can provide up to 32 data
channels (upper and lower sidebands). The channel selection for the
output data stream is controlled by the DBBC. Currently, there are two
channel mappings available - the standard geodetic format for S/X or
the ``VLBA" format which records the upper and lower sidebands of 8
channels. The data are written out at 2-bit resolution. With the
selectable IFs and bandpass filters, all
standard S/X geodetic observing modes are supported as well as those used for
astronomical and astrometric observations.
\end{sloppypar}

The VLBI Data Interchange (VSI) format data stream from the DBBC is
recorded using Mark5B+ recorders located at each site. A selectable
bit-mask can be applied to the data stream to select a subset of
channels for recording, and also the sampling quantisation. The
Mark5B+ can support recording rates of up to 1~Gbps using a single
Mark5 data module. The majority of experiments made using the AuScope
antennas have been carried out at 256 Mbps (16 8~MHz channels with
1-bit quantisation) although 1~Gbps data rates (16 16~MHz channels
with 2-bit quantisation) have been used successfully in some
experiments.

The Mark5B+ recorder is responsible for the formatting of the recorded
data. During the initialisation of the recorder, the Mark5B+ is
synchronised to the 1PPS from the DBBC via the VSI
interface. Thereafter, the 1PPS generator internal to the Mark5B+ is
responsible for the timing, using the 32 MHz VSI clock as a
reference. The Mark5B+ maintains an internal comparison between the
input and synthesised 1PPS which can be checked via the control
software. An output 1PPS is also synthesised by the Mark5B+ unit, and
the offset between this and the station GPS receiver 1PPS is used as
another way of measuring the timing stability of the system. The
Mark5B+ is synchronised to the station's GPS-Clock for establishing
the recording epoch.

\subsection{Control and monitoring systems}
\label{control_monitor}

Calibration signals, cross-over switches and IF Unit attenuation
settings are all remotely controllable through a serial
interface. This interface also permits monitoring of temperatures and
voltages throughout the system.

All of the IF and
recording systems are controlled by the PC Field System (PCFS)
computer. 
The AuScope antennas use a standard PCFS configuration (the current
9.10.04 version) with customised modules for antenna control and
system monitoring. The PCFS host machines are server-class machines
using RAID file systems for reliability.

Control and monitoring of the experiments is carried out using the
eRemoteControl interface for the PCFS, together with the Open-Monica
system \citep{monica}. eRemoteControl was developed by \cite{2010evn..confE..24N} and
offers significant benefits in bandwidth usage and connection
reliability for the remote sites. The Monica system collects
information on the observing system from a number of monitoring
points, including supply voltages for the receiver electronics,
temperature and humidity in the antenna structure, wind conditions,
drive parameters and important physical parameters such as external
temperature, atmospheric pressure and relative humidity. Most of the
analogue interfaces are provided by PICAXE-based devices which are
interfaced to Monica via simple TCP servers.

All of the information collected by Monica is permanently logged to
assist in post-facto fault finding. A real-time monitoring system is
also present to detect any faults when they occur, and to warn the
operator.

\section{The Correlator}
\label{correlator}

Following the observations, data from each station are sent to a
correlator for processing. The correlator essentially replays the
observation, combining telescope signals to provide the fundamental
measurement of geodetic VLBI: the time delays in signal arrival
between antennas on a baseline-by-baseline basis. Until recently, the
need to process the large volumes of raw data from observatories has
required purpose-built hardware correlators. However, the advent of
fast, inexpensive multi-core computer processors and high bandwidth
networking has permitted the development of flexible and scalable
software-based correlators such as  the DiFX correlator
\citep{deller2007, deller2011}.

\begin{sloppypar}
The correlator runs on the AuScope-funded Curtin
University Parallel Processor for Astronomy (CUPPA).
The correlator hardware consists of a 20 node beowulf cluster, currently with
approximately 90\,TB of attached storage, installed at, and operated
by, Curtin University in Perth, Western Australia.  The individual nodes each contain dual quad-core
2.66\,GHz Intel Xeon CPUs, 8 GB of RAM and have Gigabit ethernet
connectivity.  The correlator is connected by a 10 Gigabit ethernet
connection to iVEC (the Western Australian supercomputing centre)
allowing transfer of data from remote stations via the internet
without the necessity of shipping disks.
\end{sloppypar}

The correlator software is the latest release version of DiFX, run under a locally developed
data management software system. Data quality is checked using a
pipeline system, also developed at Curtin University, which utilises
the ParselTongue \citep{kettenis06} interface to NRAO's AIPS radio
astronomy data analysis software. It is also possible to export the
data to HOPS (Haystack Observatory Postprocessing System) for further
quality assessment and data reduction.

This combination of hardware and software is capable of processing a
standard AuScope experiment (up to 4 stations, each recording at
256\,Mbps) in approximately 30\% of the observation time.

DiFX is rapidly becoming the standard correlator for geodetic VLBI. It
is used exclusively at Curtin, at the VLBA correlator in Socorro and at the Bonn software correlator. It is also used
for geodetic VLBI processing at the IVS correlation facilities at Washington and Haystack. All of these facilities are actively
involved in the development of DiFX for geodesy and astronomy. This
active development means that it is compatible with a wide range of
back-ends and recording systems. 

DiFX has been tested against a trusted geodetic correlator
\citep{tingay2009}. Tests have been developed to ensure consistency
between versions of the correlator \citep{deller2011}. Further
verification of consistency of Geodetic and Astrometric observables is
underway (Morgan and Petrov in preparation). DiFX is fully scalable. As an example, the current CUPPA cluster is more than
capable of correlating a 10-station 4\,Gbps experiment (albeit slower
than real time).

\section{Observations and Operations}
\label{obsops}
Construction of the first AuScope telescope at Hobart was completed in
2009 and officially opened at the IVS General Meeting on 9 February
2010. Following a period of commissioning, testing and debugging, the
Hobart telescope made its first successful IVS observation in October
2010. Construction and commissioning at the other two sites continued
in parallel. Yarragadee made its first successful IVS observation in
May 2011 and, following a successful full-network fringe check on 8
June 2011, correlated at Curtin University, all three telescopes
participated in an IVS observation for the first time on 16 June 2011.

To keep operating costs at a
minimum, all three observatories were designed and constructed to be remotely
controlled and monitored. Operation of the AuScope VLBI array is being carried out from
a dedicated operations room on the Sandy Bay campus of the University
of Tasmania. Typically a single operator will control and monitor the
observations with a local person at each of the sites available
on-call when manual intervention is required. At the end of a session,
recorded media are sent to the designated correlation facility by
courier in the case of Yarragadee and Katherine but at Hobart the data
are transferred electronically via a fast network connection.

At present, the AuScope VLBI facility has sufficient operational funds
for $\sim$70 observing days per year, usually consisting of two
AuScope telescopes observing as part of the IVS network. Unfortunately
operational funds are not presently sufficient to support correlation
at Curtin University, so all data are sent to the IVS correlation
facilities at Bonn (Germany), Washington (USA) or Haystack (USA).

\section{Initial Results}
\label{results}

\subsection{Antenna Performance}

\begin{figure*}
\rotatebox{270}{\includegraphics[width=0.6\textwidth]{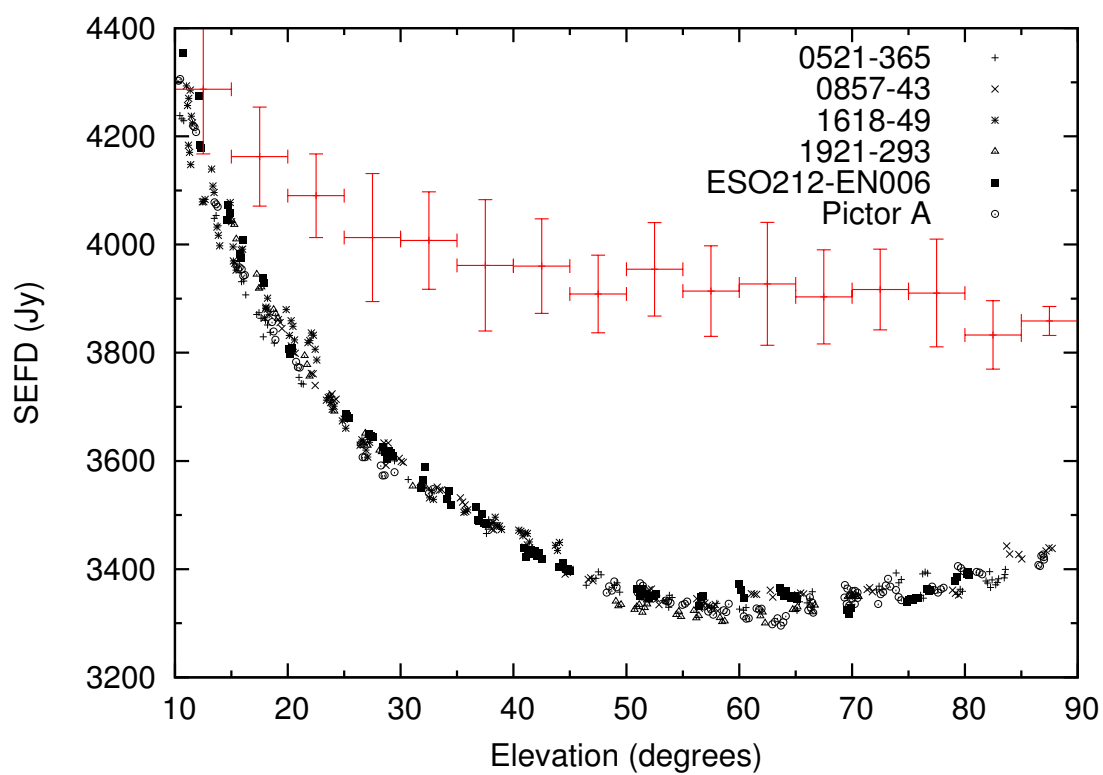}}
\caption{Measured Hobart 12\,m System Equivalent Flux Density at both S-band (red crosses) and X-band (black points) as a function of elevation. The S-band measurements have been averaged by bin, as they are somewhat noisier due to terrestrial interference.}
\label{fig:sefdx}
\end{figure*}

During the commissioning of the antennas and after any work on the
receiver system, the system temperature is measured by comparing total
output power levels with and without a microwave absorber placed over
the feed horn.  Assuming an LNA effective temperature of 30 K at
S-band and 50 K at X-band, the inferred system temperature is
generally in the range of 85--90\,K for a system in good order. The
system temperature at the Hobart telescope is currently slightly
elevated at S-band (95 K). The exact cause of this increase is under
investigation but a slight misalignment in the waveguide is
suspected. The SEFD of the telescope was estimated using sources from
the \cite{Ott94} catalogue as flux density calibrators, primarily
Virgo A and Hydra A. The zenith SEFD of the AuScope antennas is
estimated at $\sim$3500 Jy for both S- and X-band, in good agreement
with the predicted values. Measurements of SEFD as a function of
elevation at S- and X-band are shown in Figure~\ref{fig:sefdx}.

\begin{figure*}
\rotatebox{270}{\includegraphics[width=0.6\textwidth]{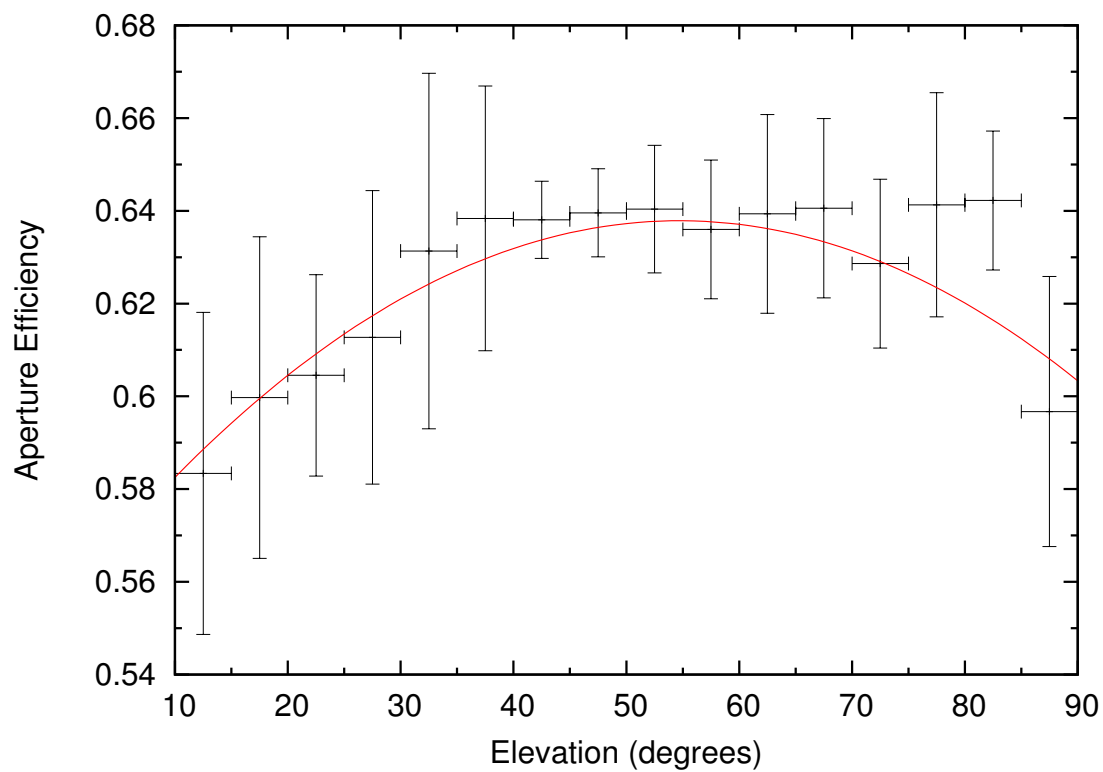}}
\caption{Hobart 12\,m aperture efficiency at X-band as a function of elevation}
\label{fig:xeffic}
\end{figure*}

The gain of the telescope was measured using observations of sources
that transited near to the zenith. The amplitude of the sources
relative to the noise diode was measured at elevations between 10 and
85 degrees. At S-band, there is no evidence for any change with
respect to elevation with an estimated aperture efficiency of 60\%. At
X-band, the optimal gain is seen at an elevation of 55 degrees and a
slight decrease is apparent toward the zenith and horizon. The peak
aperture efficiency is 64\%, decreasing to 60\% at the zenith. The
estimated values of the system temperature and aperture efficiency are
in good agreement with the predicted values for both S- and X-band.
The X-band aperture efficiency as a function of elevation (in degrees)
was estimated via a polynomial fit (Figure~\ref{fig:xeffic}) which is:
\begin{equation}
A(\mbox{El}) =  -2.77 \times 10^{-5}  \mbox{El}^2 + 3.03 \times 10^{-3}  \mbox{El} + 0.555
\end{equation}

The pointing model is currently implemented through the drive
controller itself, which accounts for structural offsets such as tilts
and encoder offsets. The RMS pointing accuracy across the sky is
estimated at 45 arcseconds. The effect of these pointing errors should
be negligible at S-band, and appear as a gain loss of $\leq 1\%$ at
X-band.

Lastly, we note that the performance in terms of telescope sensitivity
and pointing is the same at the Katherine and Yarragadee sites.

\subsection{Baseline lengths and antenna positions}

Global VLBI observations as part of the IVS were used to construct baseline length and antenna position time series.
The VLBI data were processed using the OCCAM software which adopts the least squares collocation technique \citep{titov2004}. Calculations were performed with respect to the ITRF2005 frame \citep{altamimi2007}, enabling comparisons with independent GPS positions. The Vienna Mapping Function \citep[VMF1,][]{bohm2006} was implemented, and the analysis carried out according to the IERS 2010 Conventions \citep{petit2010} with respect to Earth orientation, solid tide, ocean loading, pole tide (including the mean pole model) and thermal expansion. The only exception is for tidal atmospheric loading which has not been included. The celestial reference frame was fixed by the ICRF2 radio source positions with the exception of 39 special handling radio sources \citep{fey2009}. The geometric delay and gravitational delay models are also based on IERS 2010 Conventions \citep[see][for details]{1990SvA....34....5K,eubanks1991,1991AJ....101.2306S}.

A total of 82 daily sessions that included the Hobart~12~m (Hb) antenna were used for this analysis. These sessions comprised IVS Rapid (R1 and R4) experiments which are designed for determination of EOPs, and regional AuScope three-station experiments, scheduled for the purposes of densifying the time-series of solutions for the AuScope antennas and for monitoring and improving the catalogue of southern hemisphere reference radio sources.
For R1 and R4 sessions, the coordinates of well-established stations were used to impose the No Net Rotation (NNR) and No Net Translation (NNT) constraints to separate the EOP and geodetic parameters. The positions of the new (AuScope) stations were not constrained. However, it was found that this analysis mode was inappropriate for the regional AuScope sessions. Therefore, for those sessions we decided to first tie the Hb position to ITRF2005, and then add this station position to the list of constraints. During the first run, the Hb position was not included in the NNR and NNT constraints. After preliminary estimation of the Cartesian components, we made a second adjustment imposing NNR and NTT constraints for Hb. The time series for Katherine (Ke) and Yarragadee (Yg) comprised 24 and 14 points respectively. Due to small number statistics, we did not impose the NNR and NNT constraints for these stations.

\subsubsection{Baselines and systematic errors}

\begin{figure*}
\rotatebox{0}{\includegraphics[width=0.9\textwidth]{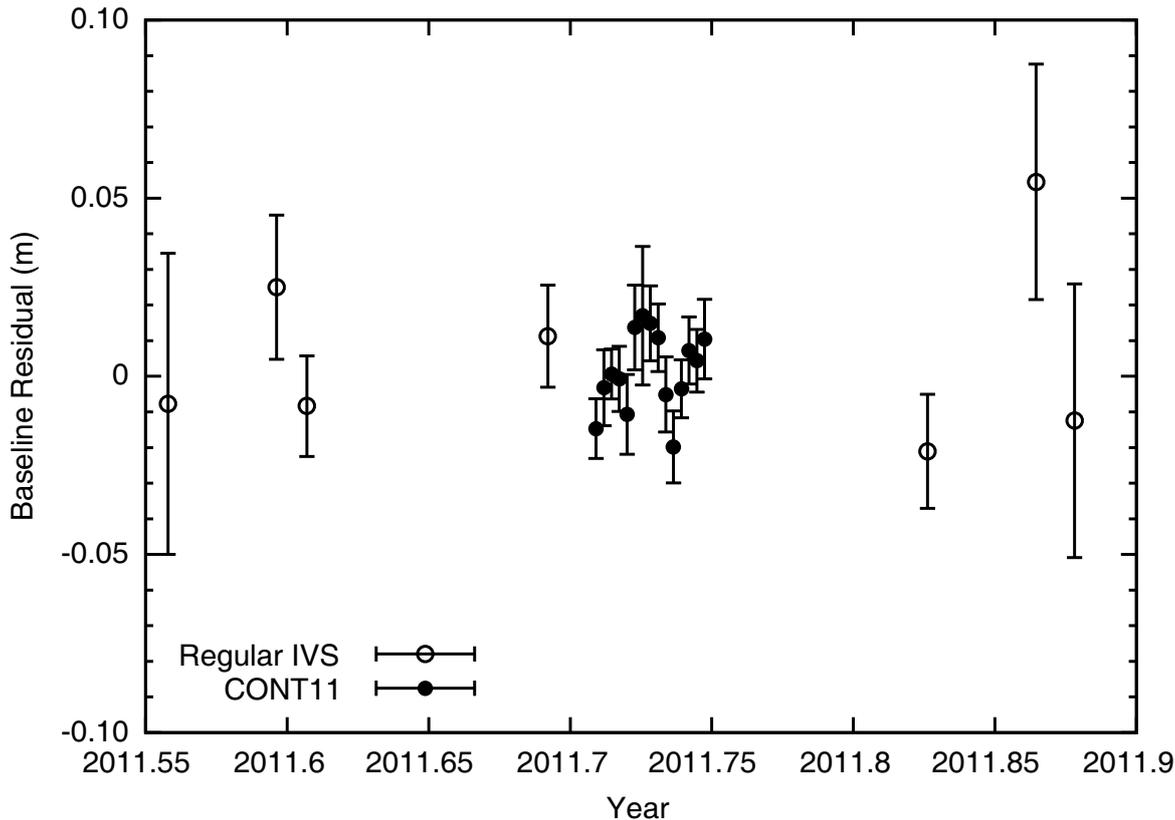}}
\caption{Time series for residuals for the Hobart 12~m -- Kokee (Hb -- Kk) baseline. Filled circles represent data from the CONT11 campaign. Open  circles are standard IVS sessions.}
\label{fig:Hb_Kk_cont11}
\end{figure*}

Baseline length repeatability is a fundamental measure of the precision of the VLBI technique. In Figure~\ref{fig:Hb_Kk_cont11} we show 
residuals for a typical long baseline between Hb and Kokee Park, Hawaii (Kk) ($\sim 8300$~km). Kokee Park is a reliable VLBI station with a large number of common observations with Hobart. With reference to Figure~\ref{fig:Hb_Kk_cont11}, the two sets of points represent different observing campaigns. Filled points represent the CONT11 campaign, an intensive 15--day set of observations involving the same 14 stations at double the usual recording rate (2 bit sampling rather than 1 bit, giving a 38\% improvement in signal-to-noise \citep{1995ASPC...82.....Z}). This is the most comprehensive and homogeneous (in terms of observing strategy) data set available. Open points are standard IVS observations (mostly the Rapid sessions), made with a range of different networks and schedules.

\begin{figure*}
 \rotatebox{0}{\includegraphics[width=0.9\textwidth]{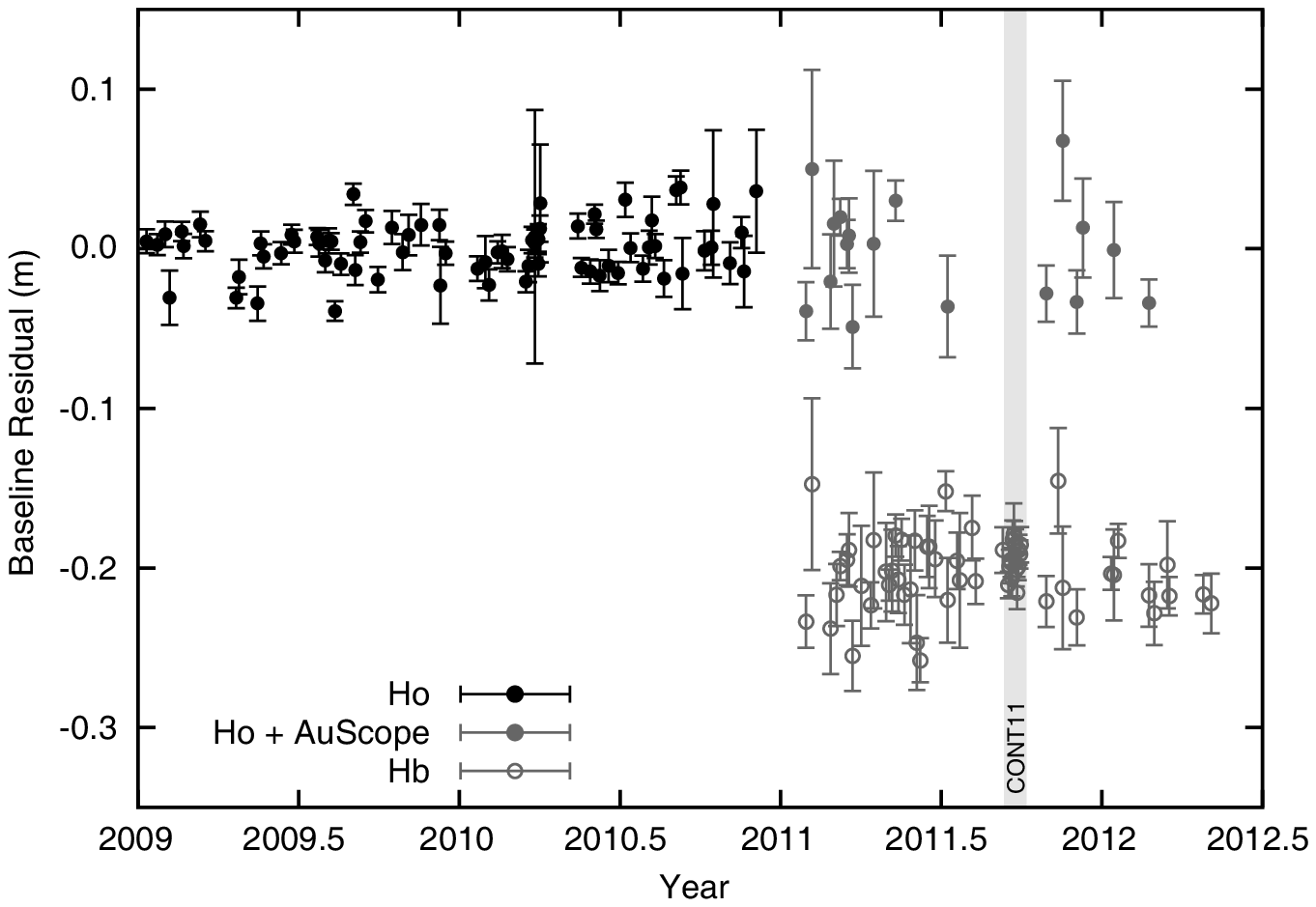}}
\caption{Time series for residuals for the baselines Hobart 26~m -- Kokee (Ho -- Kk), and Hobart 12~m -- Kokee (Hb -- Kk). Black filled circles represent data for the Ho -- Kk baseline before construction of AuScope antennas. Grey filled points show the Ho-- Kk baseline on days when at least one AuScope station also participated. Grey open points show the Hb -- Kk baseline, offset by 20 centimetres for clarity. The grey shaded region corresponds to the CONT11 campaign. With the construction and operation of the AuScope array, Ho has only been observing once per month on average in 2011.}
\label{fig:HoHb_Kk}
\end{figure*}

\begin{sloppypar}
The weighted RMS (WRMS) for the Hb -- Kk baseline length during the CONT11 campaign is 0.010~m; this is also the value of the median uncertainty in an individual 24~h measurement. This suggests centimetre accuracy in position measurements is achievable even with present techniques. By comparison, in Figure~\ref{fig:HoHb_Kk} we show the same baseline residuals for all available observations (open points). The WRMS is 0.021~m, and median formal uncertainty 0.017~m.
\end{sloppypar}

Our results highlight an important point. While formal uncertainties in station positions and baseline lengths can be reduced by increasing the number of observations (since $\sigma \propto N^{-1/2}$), this is not particularly useful when attempting to assess the quality of an individual position measurement. It is clear from Figures~\ref{fig:Hb_Kk_cont11} and \ref{fig:HoHb_Kk} that the main factor at present limiting the accuracy of VLBI observations lies in systematic biases inherent to the analysis. For example, ignorance of quasar structure and variability (which can exhibit different temporal behaviour at S and X bands; e.g.  \citealt{shabala2012}) will map into source and station positions. Furthermore, the magnitude and sign of these effects will in fact depend in a complicated fashion on network geometry and the observing schedule. Charlot and collaborators \citep[e.g.][]{charlot1990,feyandcharlot1997,tornatorecharlot2007} have shown that group delays due to quasar structure of as much as 1 nanosecond (corresponding to a 30~cm position shift) are possible in individual quasars. 

Using the structure index \citep{feyandcharlot1997} distribution of ICRF2 quasars \citep[see Figure 12 of][]{2009ITN....35....1M} and the typical number of observations for each of these quasars, \citet{shabala2013} have calculated the uncertainty in the average group delay due to source structure. These authors find that this value corresponds to position uncertainties of 2.4~cm in the southern hemisphere, and 1.6~cm in the northern hemisphere. They note that the exact value is a strong function of the quasar selection strategy. The distribution of antennas in a network, and time of observation, are important since it is the source structure projected onto a given baseline that determines the additional time delay. Detailed simulations of IVS observations incorporating quasar structure will help address this issue more quantitatively.

We illustrate the significance of systematic errors in Table~\ref{tab:wrms_vs_sigma}, where we assess the statistical significance of observed weighted RMS (WRMS) for each set of measurements presented in this Section. We have run Monte-Carlo simulations for each set of observations, and calculated the probability of obtaining the observed WRMS by chance. For Hb -- Kk observations made in the ``normal' mode, the WRMS is significantly different (at the 95 percent level) from the observed formal errors. In other words, the scatter is too large to be due to random scatter alone. On the other hand, the WRMS and formal uncertainties are consistent with each other for the same baseline observed as part of the CONT11 campaign. 

\begin{table*}
\caption{Baseline length repeatability. Observed scatter (WRMS), formal uncertainties ($\sigma$) in baseline measurements, and number of observations are shown in the first three columns. WRMS expected from Monte-Carlo simulations, and probability of observed WRMS being consistent with formal uncertainties are shown in the last two columns.}
\label{tab:wrms_vs_sigma}
\begin{tabular}{lccccc}
\hline
Baseline & $\sigma$ (m) & observed  & No. sessions & expected & prob(WRMS | $\sigma$) \\
&  & WRMS (m) &  & WRMS (m) & \\\hline\hline
Hobart26 (Ho) -- Kokee (Kk), no AuScope	& 0.012 & 0.016 & 75 & 0.010 & 0.0001\\
Hobart26 (Ho) -- Kokee (Kk), with AuScope	& 0.028 & 0.031 & 17 & 0.025 & 0.097\\\hline
Hobart12 (Hb) -- Kokee (Kk), no CONT11		& 0.022 & 0.0238 & 45 & 0.0201 & 0.064\\
Hobart12 (Hb) -- Kokee (Kk), CONT11 only	& 0.0104 & 0.0107 &15 & 0.0100 & 0.348\\
Hobart12 (Hb) -- Hobart26 (Ho) 			& 0.0081 & 0.0076 & 22 & 0.0073 & 0.381\\
\hline
\end{tabular}
\end{table*}

We note that the WRMS on the Hb -- Kk baseline halves from 0.021~m during regular IVS sessions to 0.010~m for the 15 days of CONT11 observations. These observations were made with an array of the same 14 antennas and the same source catalogue for every session. A marked improvement in baseline length repeatability is achieved, despite the fact that different observing schedules were used for each session, potentially introducing different atmospheric and quasar structure biases into solutions (although the number of observations of a given source per session was roughly constant). This result indicates the important role that network geometry also plays in systematic errors and emphasises the need for further simulations and observations to disentangle and quantify the error contribution from these two effects.

Also shown in Figure~\ref{fig:HoHb_Kk} is the Hobart26 (Ho) -- Kokee Park baseline, before (black points) and after (grey filled points) the construction of the AuScope array. Inclusion of AuScope antennas is expected to improve both the formal errors on individual sessions, and repeatability (WRMS) of the Ho -- Kk baseline. We obtain ($\sigma$,WRMS) = (0.012,0.016)\,m for the period 2009 -- 2010 when no AuScope antennas were involved; and (0.028,0.031)\,m in 2011--2012, when AuScope antennas are included. The pre-AuScope RMS is inconsistent with $\sigma$ at the 99.9 percent level (Table~\ref{tab:wrms_vs_sigma}) again suggesting systematic sources of position error. Due to higher formal errors, the post-AuScope WRMS and $\sigma$ do not differ in a statistically significant way. The formal errors increase due to Ho being scheduled for a much lower number (a factor of $\sim 2$) of observations when Hb is included in the schedule. Comparing the velocities of the Ho -- Kk and Hb -- Kk baselines, we get $0.047 \pm 0.003$ and $0.059 \pm 0.009\, \mbox{m} \mbox{a}^{-1}$ respectively. The velocities are consistent within the measured uncertainties. The larger formal baseline errors when AuScope antennas are included are consistent with results for individual antenna positions (Tables~\ref{tab:posGeo} and \ref{tab:posXYZ}). This, in turn, is caused by teething problems during early commissioning at the stations, such as for example frequent clock discontinuities. Another, more subtle, effect is network geometry. AuScope antennas observe in more Southern Hemisphere--dominated networks than did the Hobart 26\,m. The scarcity of short baselines means there are very few sources at high mutual elevation; this affects the quality of atmospheric solutions. Furthermore, quasars at low declinations (below $-30^\circ$) are much less well studied than their Northern Hemisphere counterparts. Many of them show highly variable positions \citep[see e.g. Figure~33 of][] {2009ITN....35....1M}, presumably due to source structure. These effects all contribute to degrading solutions involving  the AuScope antennas.

\begin{figure*}
\rotatebox{0}{\includegraphics[width=0.9\textwidth]{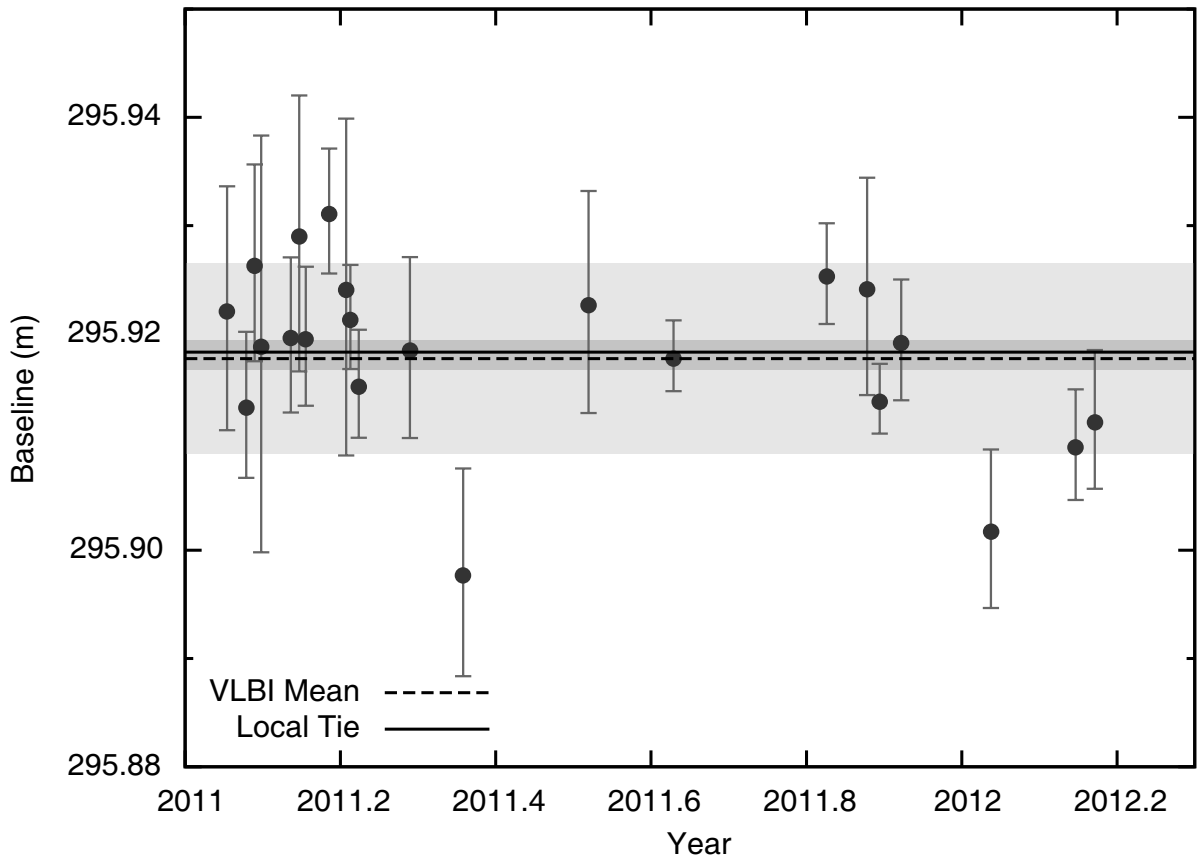}}
\caption{Time series for the Ho-Hb baseline. The mean baseline length is shown by the dashed line. Results of the local tie survey are shown by the solid black line. Shaded regions are WRMS on the VLBI mean (light grey), and uncertainty in local tie measurement (dark grey).}
\label{fig:Ho_Hb}
\end{figure*}

In Figure~\ref{fig:Ho_Hb} we show the evolution of the Hb-Ho baseline, noting that both antennas are located on the same bedrock, separated by $\sim 296$\,m. The mean baseline length is 295.917\,m, with ($\sigma$,WRMS) = (0.008,0.007)\,m. These residuals are due to effects such as thermal and gravitational deformation of the antennas, clock stability, and structure of quasars making up the ICRF. Clock stability is an issue here because Ho and Hb currently run off independent time and frequency standards and source structure plays a role as the Ho--Hb baseline measurements were made as part of a global solution. The measured baseline length is in agreement with the results of the Mt.\,Pleasant local tie survey \citep{Rudandwoods} of $295.918 \pm 0.001$\,m. It is interesting that individual station positions (Table~\ref{tab:posXYZ}) agree less well with the local tie survey than the baseline length. This is consistent with errors in individual solution components mapping into estimates of other components \citep{nothnagel2002}.

\subsubsection{Antenna positions}

Station positions derived from these observations are presented in Tables~\ref{tab:posGeo} and \ref{tab:posXYZ}. Also shown are the station velocities for the Hobart antenna. In Table~\ref{tab:posGeo} the inner uncertainties represent the average RMS of individual measurements; while the outer uncertainties denote scatter in best-estimates (i.e. WRMS about the trend line). In other words, the first number corresponds to formal uncertainty given by an OCCAM solution for a typical 24\,h session. The second number indicates how repeatable the station position or velocity estimate is from one session to the next. We also quote results of the local tie survey for the Hobart antennas \citep{Rudandwoods}. In geodetic VLBI, formal uncertainties in baseline lengths and station positions are usually represented via the WRMS (our second set of uncertainties) divided by $\sqrt{N}$, where $N$ is the number of observations. However, in light of the systematic effects discussed above, which we believe dominate VLBI measurements, we refrain from quoting such overly optimistic values.

\begin{table*}
\caption{Calculated positions for the three AuScope VLBI antennas at epoch 2012.0, ITRF2005 datum. The first uncertainty represents the average formal error in an individual measurement. The second uncertainty corresponds to weighted scatter about the best estimate.}
\label{tab:posGeo}
\begin{tabular}{|l|r@{\,}l@{\,}l@{$\,\pm\,($}l@{\,}l|}
\hline
\multicolumn{6}{|c|}{Hobart 12\,m (Hb)} \\ \hline
No. Sessions & \multicolumn{5}{|l|}{82} \\
Latitude (d m s)     & $-42$ & 48 & 20.0621 & $0.0006,$ & $0.0012)$  \\
Longitude (d m s)   & 147    & 26 & 17.3070 & $0.0006,$ & $0.0012)$ \\
Height (m)  &           &     &    40.967 & $0.019,$   & $0.036)$ \\\hline\hline
\multicolumn{6}{|c|}{Katherine 12\,m (Ke)} \\ \hline
No. Sessions & \multicolumn{5}{|l|}{24} \\
Latitude (d m s)     &  $-14$ & 22 & 31.6679 & $0.0006,$ & $ 0.0004)$ \\
Longitude (d m s)  & $132$ & 09 & 08.5439 & $0.0007,$ & $ 0.0006)$ \\
Height (m) &            &      &  189.262 & $0.021,$    & $0.017)$  \\\hline\hline
\multicolumn{6}{|c|}{Yarragadee 12\,m (Yg)} \\ \hline
No. Sessions       & \multicolumn{5}{|l|}{14} \\
Latitude (d m s)    & $-29$ & 02 & 49.7226  & $0.0012,$ & $0.0006)$\\
Longitude (d m s) &  $115$ & 20 & 44.2576 & $0.0013,$ & $0.0013)$\\
Height (m)            & & & 248.236 & $0.040,$ & $0.029)$ \\\hline
\end{tabular}
\end{table*}

\begin{table*}
\caption{Calculated positions for the three AuScope VLBI antennas and Hobart26, and Hobart12 velocities at epoch 2009.9 in geodetic coordinates, ITRF2005 datum. Quoted uncertainties are WRMS values. Local tie survey results of \cite{Rudandwoods} for the Hobart antennas are also shown.}
\label{tab:posXYZ}
\begin{tabular}{|l|r@{$\,\pm\,$}l|r@{$\,\pm\,$}l|r@{$\,\pm\,$}l|}
\hline
Station & \multicolumn{2}{|c|}{X (m)} & \multicolumn{2}{|c|}{Y (m)} & \multicolumn{2}{|c|}{Z (m)}\\\hline\hline
Hobart26 (Ho) VLBI	& $-3950237.247$ & $0.013$ & $2522347.670 $ & $0.008$ & $-4311562.033$ & $0.018$\\
Hobart26 (Ho) local tie	& $-3950237.251$ & $0.0006$ & $2522347.669 $ & $ 0.0003$ & $-4311562.015 $ & $ 0.0004$\\\hline
Hobart12 (Hb) VLBI	& $-3949990.687$ & $0.044$ & $2522421.191 $ & $ 0.012$ & $-4311708.164 $ & $ 0.032$\\
Hobart12 (Hb) local tie	& $-3949990.676 $ & $ 0.0004$ & $2522421.199 $ & $ 0.0003$ & $-4311708.170 $ & $ 0.0003$\\
Hobart12 (Hb) velocity $(\mbox{m} \mbox{a}^{-1})$ & $ -0.038 $ & $ 0.009$ & $-0.001 $ & $ 0.004$ & $0.046 $ & $ 0.007$\\\hline
Katherine (Ke) VLBI			& $-4147354.453 $ & $ 0.018$	& $4581542.384 $ & $ 0.017$ & $-1573303.310 $ & $ 0.012$\\\hline
Yarragadee (Yg) VLBI		& $-2388896.040 $ & $ 0.036$ & $5043349.973 $ & $ 0.029$ & $-3078590.989 $ & $ 0.015$\\
\hline
\end{tabular}
\end{table*}

\section{Conclusions}

We have constructed, commissioned and are now operating three new 12\,m radio telescopes for geodetic VLBI on the Australian continent. We have also constructed and tested a new software correlator facility for geodesy at Curtin University. For the observatories, we have followed a design strategy to match the VLBI2010 requirements as closely as possible and to allow an upgrade to full VLBI2010 capabilities in the future. All three telescopes are operated remotely with a high level of reliability. The individual telescopes are meeting their design specifications in terms of mechanical performance and sensitivity. As elements of a VLBI array we have demonstrated their capability to achieve centimetre precision in typical baseline measurements and have presented non-variant point positions for all three telescopes.

Data taken during the 15 day CONT11 campaign have shown the dramatic improvement to positional accuracy and repeatability that can be achieved by observations with a non-varying global array of telescopes. We suggest that further reductions in systematic errors can be made by careful selection of radio sources and by comparing observations made using suitable quasars with different schedules. Investigation of these effects is being undertaken through the AUSTRAL observing program within the IVS which is using the three AuScope telescopes and the near-identical 12\,m telescope at Warkworth, New Zealand \citep{nz}. These observations are being undertaken to strengthen the Southern Hemisphere reference frame and to exercise the full capabilities of these small, fast antennas with high data recording rates, which allow for typically twice the number of observations per telescope per day as the IVS Rapid observations. The improved observation rate, constant array configuration, judicious source selection and scheduling is expected to further reduce systematic errors. We are also using surveying techniques to study the thermally and gravitationally induced deformation of the 12\,m telescopes in order to quantify and subsequently mitigate this systematic effect.

The most important upgrade now required to the AuScope VLBI array is a $\sim 2 - 14$~GHz broadband receiving and recording system to bring it significantly closer to the ultimate VLBI2010 aim of millimetre accurate positions.

\begin{acknowledgements}
This study made use of data collected through the AuScope initiative. AuScope Ltd is funded under the National Collaborative Research Infrastructure Strategy (NCRIS), an Australian Commonwealth Government Programme. JM ad SS were supported under the Australian Research Council's Super Science funding scheme (projects FS110200045 and FS100100037). We wish to thank our Reviewers for a comprehensive and constructive list of comments. Lastly we wish to thank the International VLBI Service for Geodesy and Astrometry (IVS) for access to VLBI data products used in this paper and for advice and support during the conception, design, commissioning and operation of the AuScope VLBI Array.
\end{acknowledgements}

\bibliographystyle{natlib}      
\bibliography{Lovell_etal}

\end{document}